\documentclass[epjST]{svjour}
\usepackage{mathptmx}
\usepackage{helvet}
\usepackage{tabularx}
\usepackage{courier}
\usepackage{pifont}
\usepackage{cite}
\usepackage{todonotes}
\usepackage{floatflt}
\usepackage{tocloft,hyperref}

\begin{document}

\author{Norbert Marwan\inst{1}\fnmsep\thanks{\email{marwan@pik-potsdam.de}}}
\institute{Potsdam Institute for Climat Impact Research, 14412 Potsdam, Germany}

\title{Historical Review of Recurrence Plots}

\abstract{
In the last two decades recurrence plots (RPs) were introduced in many different scientific disciplines. It turned out how powerful this method is. After introducing approaches of quantification of RPs and by
the study of relationships between RPs and fundamental properties of dynamical systems, this method attracted even more attention. After 20 years of RPs it is time to summarise this development in a historical context.
}
\maketitle

\section{Introduction}
\label{sec1}

The technique known as recurrence plot is 20 years
old. However, recurrences were studied and employed
long before. The Maya calender\index{Maya calender} is one example where 
we can find the principle of recurrences as the basic
idea. We encounter recurrences in different
aspects in nature and social life.

With the birth of the modern mathematics
in the 19th century recurrence was discovered
to be a fundamental property of conservative dynamical systems.
Poincar\'e\index{Poincar\'e} formulated his thesis in the work about 
the restricted three-body system, which won him a prize
sponsored by Oscar II of Sweden and Norway\index{Oscar II of Sweden and Norway}. Poincar\'e
found that ``In this case, neglecting some exceptional
trajectories, the occurrence of which is infinitely 
improbable, it can be shown, that the system recurs 
infinitely many times as close as one wishes to its 
initial state.''\index{Poincar\'e recurrence} (translated from \cite{poincare1890}).
In the following years, several important mathematical works 
were performed (e.g.~\cite{kac47}).

However, more than a half century had to pass for 
recurrences to be comprehensively studied on 
numerical simulations and real measurements. Not until
the introduction of powerful computers
such numerically costly studies were possible. As an 
example, we may take the Lorenz system\index{Lorenz system}, which 
was one of the first numerical models exhibiting recurrences
and chaotic behaviour \cite{lorenz63}. Recurrences
were analysed by {\it first return maps}\index{first return maps} \cite{procaccia1987},
{\it space time separation plots}\index{space time separation plots} \cite{provenzale92},
{\it return time}\index{return time statistics} and {\it recurrence time statistics}\index{recurrence time statistics} 
\cite{balakrishnan2000,hirata99}

The persistent growth of computer power allowed 
even more computer intense investigations, as a
pair-wise comparison of all possible combinations of
pairs of a data series. This can be done by the similarity  
matrix, a graphical representation of the similarity
of all pair-wise combinations in the considered
data series. Although strictly speaking, the idea 
of a distance metric\index{distance metric} can be traced back to the
Pythagorean Theorem\index{Pythagorean Theorem}, the modern concept of this tool dates back
to the 1920s in both applicative \cite{thurstone1927}
as well as methodological fields \cite{mahalanobis1936}.
The work of Kruskal\index{Kruskal} in the 1960s \cite{kruskal1964a} 
was one of the most quoted works 
in statistics and deeply affected many fields of 
investigation from ecology to psychology and economics. 
All these fields appeared as separate by 
physical science so that the appreciation of these works 
remained limited in physics. However, these authors deeply 
investigated and exploited this approach for
an analysis of distance spaces allowing for an unbiased 
representation of virtually all kind of data without 
any constraint about their characteristics. In this manner, they
paved the way for the nowadays recognized ability of recurrence 
based methods to deal with non-stationary, non-linear and 
relatively short data series. 

With the intense usage of
computers, the similarity matrix was re-invented
by several scientific disciplines 
around the change from the 1970s to 1980s, and therefore
different terms for the same technique,
like {\it dot plot}\index{dot plot} \cite{church1993}, 
{\it contact map}\index{contact map} \cite{domany2000,holm1993}, 
{\it similarity matrix}\index{similarity matrix} \cite{krishnan1986,kruskal1983}
or {\it distance matrix}\index{distance matrix} \cite{sakoe1978} emerged.
In the field of chaos theory it found its way a few
years later as the {\it recurrence plot}\index{recurrence plot}
\cite{eckmann87} (Fig.~\ref{cl_ret_plot}A). Now the aim was to compare
all possible states represented by a higher-dimensional
phase space trajectory. In case the trajectory runs
through a region in the phase space it passed before,
then we consider it as a recurrence. A recurrence
means that the recurrent state is somehow similar
to a former state. This definition of similarity
offers leeway to adopt the method to the needs of
the investigation, as we will see later. Thus, the
recurrence plot technique was not really new.
The intention of Eckmann et al.~was to have
another representation of the dynamics of the
systems. However, they immediately noted that further
important information, like determinism, divergence
and drifting behaviour can be found in such plots.
They also stated that the lengths of the diagonal
line structures in the RP are related to the
positive Lyapunov exponent.

\section{The birth of the recurrence plot}
\label{sec2}

By utilisation of the similarity matrix as a
tool to visualise recurrences of higher-dimensional
phase space trajectories, Eckmann et al.~did
not expect to establish a new direction in
nonlinear data analysis. Nevertheless, 1987 is
considered to be the birth of {\it recurrence plots}
and their quantification as a modern tool of
nonlinear data analysis.\index{recurrence plot|birth}

\begin{figure}
\centering \includegraphics[width=.9\columnwidth]{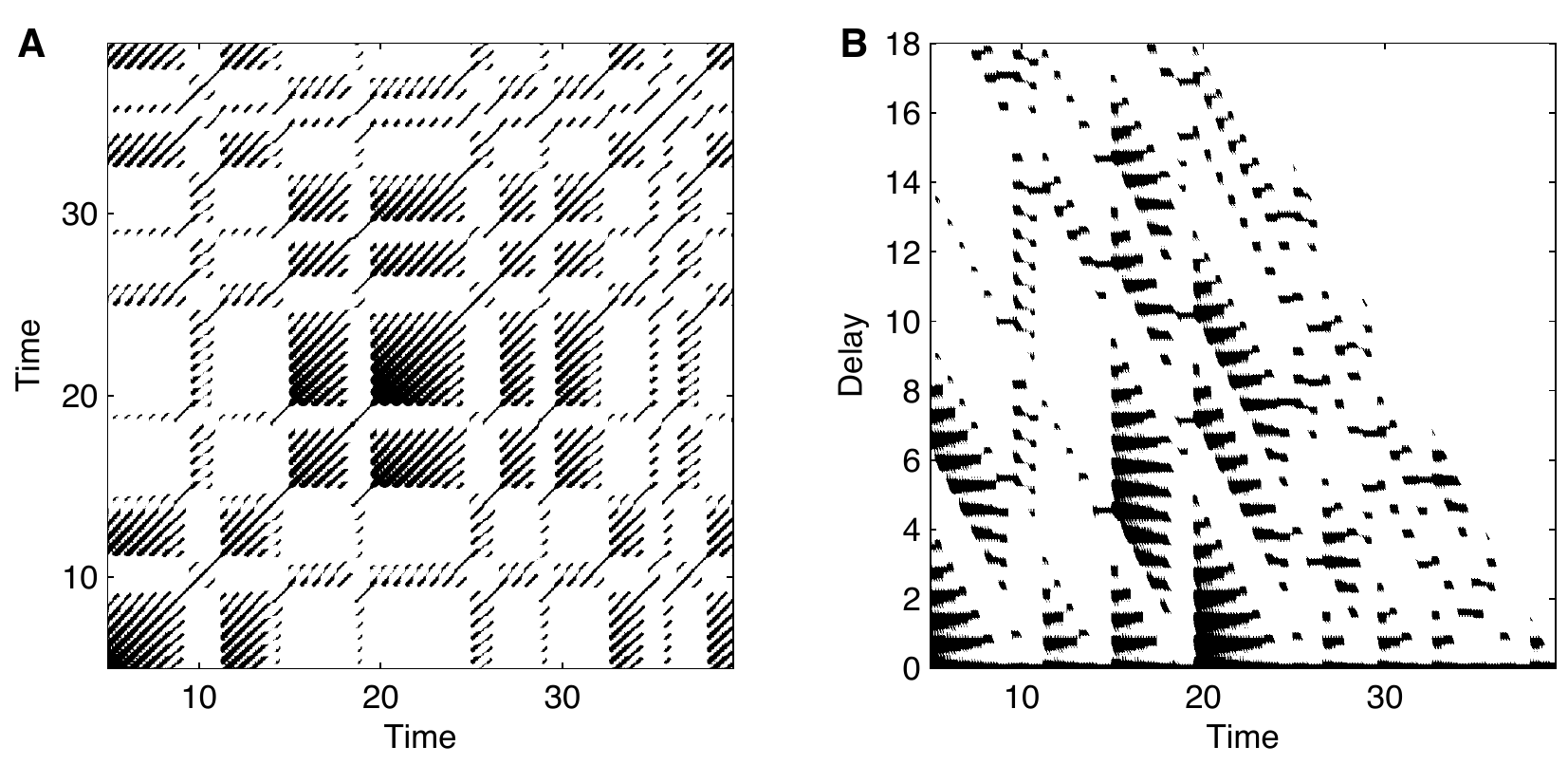}
\caption{(A) Recurrence plot and (B) ``close returns plot''
of the $x$-component of the Lorenz system \cite{lorenz63}.
Used RP parameters: $m=5$, $\tau=5$, $\varepsilon=7.6$, 
$L_{\infty}$-norm.}
\label{cl_ret_plot}
\end{figure}

Short time later (no later than 1992), 
different authors independently
introduced another kind of representation of recurrences
\cite{mindlin92,zbilut90}. They
did not compare all possible time points, but
only a given time into the past and future (Fig.~\ref{cl_ret_plot}B). 
Here a further
name appeared: the {\it close returns plot}\index{close returns plot}.
Such a representation can be more intuitive, 
in particular for beginners, because the line
structures of the recurrence plot will be parallel to
the $x$-axis.

%
%
%
%

\section{Recurrence quantification analysis}
\label{sec3}

\begin{figure}
\centering \includegraphics[width=.65\columnwidth]{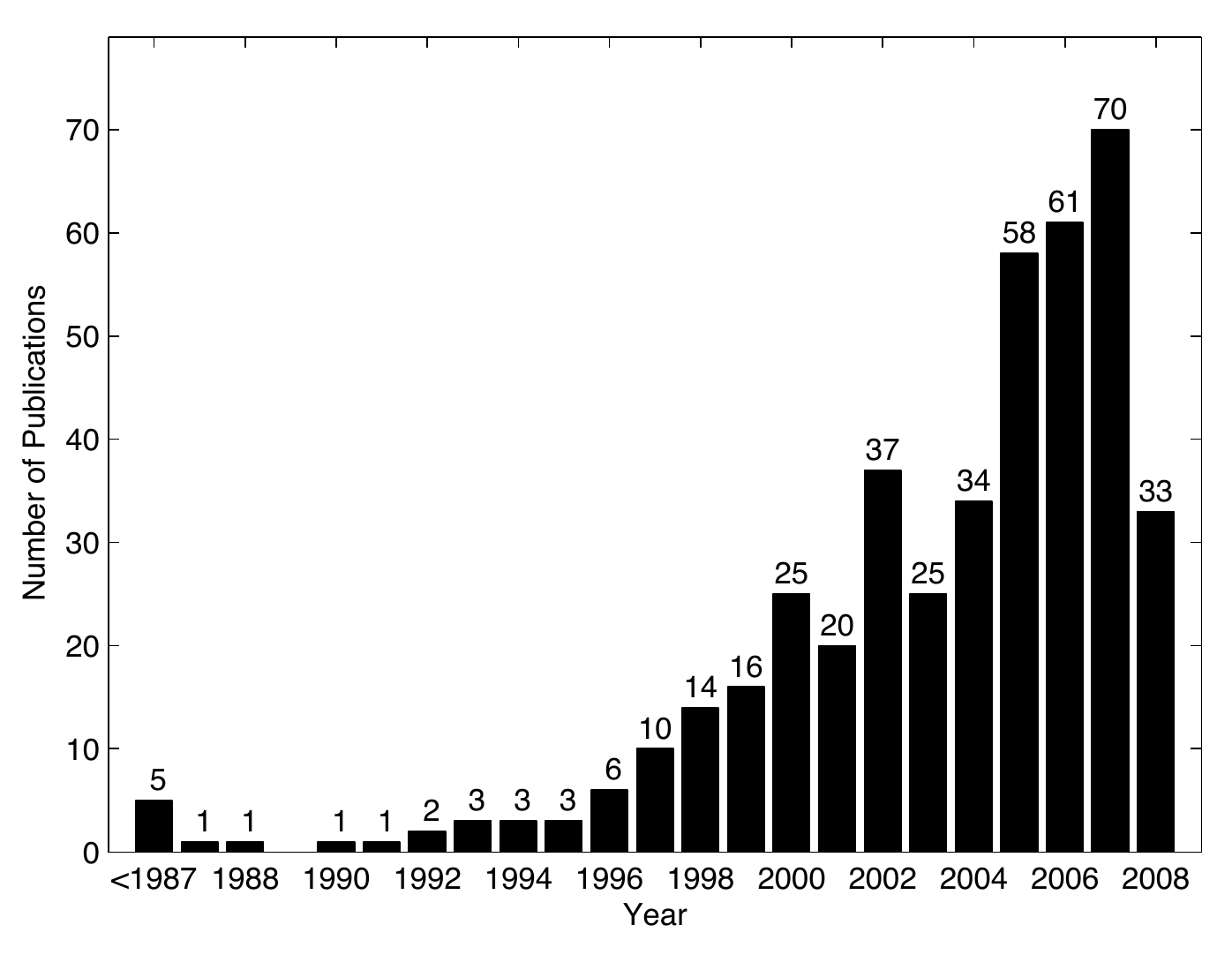}
\caption{Publications about RPs and RQA for the last 20 years (May 2008).}
\label{publications}
\end{figure}

These first years were characterised by a rather rare
application of this method (Fig.~\ref{publications}). 
The appearance of
recurrence plots in publications was somehow
exotic. Moreover, up to this time, recurrence plots
were just a visualisation tool, what yielded
to the disadvantage that the user had to detect
and interpret the patterns and structures revealed
by the recurrence plot. Low screen and printer 
resolutions further worsened this issue.
To overcome this subjective part of the method,
starting in the late eighties, Zbilut and Webber
tried to quantify the structures of the RP.\index{recurrence quantification|birth} 
At first they just determined the density of 
recurrence points in the RP and studied the 
histogram of the lengths of diagonal lines
\cite{webber94,zbilut92,zbilut2007a}.
In the following five years, they introduced
the known measures of complexity based on
diagonal line structures of recurrence plots
and therewith established the {\it recurrence quantification
analysis (RQA)}:
\begin{itemize}
\item percentage recurrences or recurrence rate\index{percentage recurrences}
\item percentage determinism\index{percentage determinism}
\item maximal line length and divergence\index{maximal line length, divergence}
\item Shannon entropy of the distribution of the line lengths\index{Shannon entropy}
\item trend\index{trend}.
\end{itemize}
For a definition of these measures we refer
to \cite{marwan2007}. The usefulness of
these measures was shown by an increasing
number of applications to real data. However,
until 1995, only few applications of RPs and
RQA appeared in publications.

Since the early nineties, Webber
provides a freely available software ({\it RQA
Software})\index{RQA
Software} which can be used to compute RPs
and the RQA measures. In 1996, Kononov started
the {\it Visual Recurrence Analysis (VRA)}\index{Visual Recurrence Analysis (VRA)} software.
It has a user-friendly graphical interface and 
computational enhancements. Therefore, this software
is rather popular. The TISEAN\index{TISEAN} package, provided 
by Hegger, Kantz and Schreiber, was also one of the
first software packages able to compute RPs (but without
quantification, just RPs). For locations of these
software in the WWW we refer to the web site
\url{http://www.recurrence-plot.tk}.

As a next milestone we find the introduction of
the time-dependent RQA\index{recurrence quantification|time-dependent}. The RQA measures are
calculated from windows moved along the
main diagonal of the RP. This allows for the study
of the evolution of the RQA measures over time
\cite{trulla96}. It was shown that with this
approach it would be possible to detect transitions\index{transitions}
in dynamical systems. At this moment, only
transitions between regular and non-regular
dynamics (like period-chaos transitions)\index{transitions|period-chaos} could
be detected. In the same year,
a publication with the promising title
``Recurrence plots revisited'' by \cite{casdagli97}
appeared. It suggested to use RPs to reconstruct the 
driving force\index{driving force} of dynamical systems and
introduced the idea of {\it meta recurrence plots}\index{meta recurrence plot},
based on windowing and correlation sum.

The major methodological work on the RP and RQA 
during the 1990s was performed by the group
around Zbilut and Webber in Chicago. 
Since the mid-1990s, the scientific community became more
and more aware of RPs, as the continuously
increasing number of
publications between 1996 and 2004 demonstrates
(Fig.~\ref{publications}).

%

Towards the end of the 1990s, first theoretical
studies on the RP regarding their relationship
with dynamical invariants and the preservation
of the topology appeared. McGuire et al.~analytically 
demonstrated that the distance
matrix as the base of the RP preserves
all information to reconstruct\index{reconstruction}
the underlying data series \cite{mcguire97}. Faure and Korn
have shown that the cumulative distribution
of the lengths of the diagonal lines is
directly related to the $K_2$ entropy\index{$K_2$ entropy} \cite{faure98}. The link
between the columns of the RP and the {\it 
information dimension}\index{information dimension} was discussed by Gao and Cai
\cite{gao2000}.

In 1998,
Iwanski et al.~already discussed the issue
whether it is really necessary to embed\index{embedding} in order 
to derive quantities for the description
of the dynamics. The authors based their 
discussion on more heuristic numerical work
and by using the RQA measure {\it maximal
line length}. This issue was further discussed
by Gao and Cai \cite{gao2000}, who also
used the RP in order to estimate recurrence
times. They defined two types of recurrence
times\index{recurrence time} based on the vertical distance between 
recurrence points in the RP. 

In 1999, the {\it perpendicular RP}\index{perpendicular recurrence plot} was suggested as
a refinement in order to estimate the divergence
of the states \cite{choi99}. Here a recurrence
is defined using the additional condition that
the recurrence points have to lie on a plane which is
perpendicular to the phase space trajectory of the
reference point. 
The {\it iso-directional RP}\index{iso-directional recurrence plot}, introduced in 2002, goes in 
a similar direction \cite{horai2002}. Its additional recurrence condition
requires that the recurrent phase space trajectories have to 
evolve in parallel, i.e.~in the same direction.
Unfortunately, these variants of a RP
are not popular, probably because of their higher computational
efforts.

\section{Extensions for the recurrence plot and quantification analysis}
\label{sec4}

Also around the change to the new millennium, the
RP technique was extended to the bivariate
{\it cross recurrence plot (CRP)}\index{cross recurrence plot} \cite{marwan99,zbilut98}.
This bivariate extension tests for simultaneous
occurrences of similar states in two different
systems. Consequently, cross recurrence 
quantification analysis\index{cross recurrence 
quantification analysis} followed. This technique
can be used to detect deterministic signals \cite{zbilut98}
and to study complex interrelations between
different systems \cite{marwan2002pla,marwan2003climdyn}. 
Here delay based variants of the RQA measures were introduced 
\cite{marwan2002pla}.
Furthermore, CRPs appeared rather illustrative to
study differences or transformations of time scales
of similar observations \cite{marwan2002npg}.
This feature was later used to understand changing 
shapes of line structures\index{recurrence plot|line structures} in RPs \cite{marwan2005}.
The detection of deterministic signals by using RQA
was further demonstrated by Zbilut et al.~ \cite{zbilut2000a}.

With the introduction of CRPs, the freely available
{\it CRP Toolbox}\index{CRP Toolbox} for MATLAB, written by Marwan, appeared.
This toolbox is platform independent and contains almost all RP related 
tools and measures. It is noteworthy that also
commercial software started to include at least the computation
of RPs, like {\it Dataplore} (ixellence GmbH, Germany).
For locations of these software in the WWW we again 
refer to the web site\index{recurrence plot|software}
\url{http://www.recurrence-plot.tk}

With the new millennium, further measures of complexity
were added to the RQA. Marwan et al.~introduced
measures based on vertical line\index{recurrence plot|vertical line}
structures in the RP and are called {\it laminarity}\index{laminarity}
and {\it trapping time}\index{trapping time} \cite{marwan2002herz}. 
Using these measures it was
possible to detect chaos-chaos transitions.\index{transitions|chaos-chaos}

At the same time, in bio-informatics RPs and RQA 
were employed to investigate the spatial structure 
of biopolymers \cite{giuliani2002a}.\index{spatial analysis} This was a deep change
in perspective, because here these methods do not analyse
time series but spatial series or even spatial structures 
(starting directly from distance matrices without the 
need of a pre-existing series, \cite{webber2001})
and makes the technique 
to come back to its `purely statistical' lineage 
(as opposite to the dynamical lineage).

\section{Theoretical basis and dynamical invariants}
\label{sec5}

Between 2002 and 2006, Romano and Thiel published several
pioneering articles related to different aspects of
RPs. They theoretically justified the choice of the
recurrence threshold\index{recurrence threshold} for data with observational
noise\index{observational noise} and were able to analytically describe
an RP for noise \cite{thiel2003,thiel2002}. They explained
the link between the line lengths of the diagonal
lines and the dynamical invariants\index{dynamical invariants} \cite{thiel2004a}.
This work led to further studies about the influence
of embedding\index{embedding|influence} \cite{march2005,thiel2004b,thiel2006}

In 2004, a real multivariate extension of RPs, the
{\it joint recurrence plot (JRP)}\index{joint recurrence plot} was introduced \cite{romano2004}.
JRPs test for simultaneous occurrences of 
recurrences in different systems and are proper means
for the detection of general synchronisation\index{synchronisation|general synchronisation} \cite{romano2004b}.
Romano et al.~have further demonstrated how to
use a delay based RQA measure for the detection of
phase synchronisation\index{synchronisation|phase synchronisation}, even for non-phase coherent
oscillators \cite{romano2005}. This technique can be used to detect the
direction of the coupling between systems\index{coupling direction} \cite{romano2007}. 
During this time, the idea of twin surrogates\index{twin surrogates} appeared,
which are dynamics preserving surrogates based on recurrences
\cite{marwan2007,thiel2008}. Such surrogates can be used
to derive a statistical inference for a synchronisation analysis.
Moreover, a spatial extension\index{recurrence plot|spatial extension} of RPs was introduced, resulting
in RPs of higher dimension (like 4D or 6D) \cite{marwan2007pla}.

%
%
\begin{figure}
\centering \includegraphics[width=.9\columnwidth]{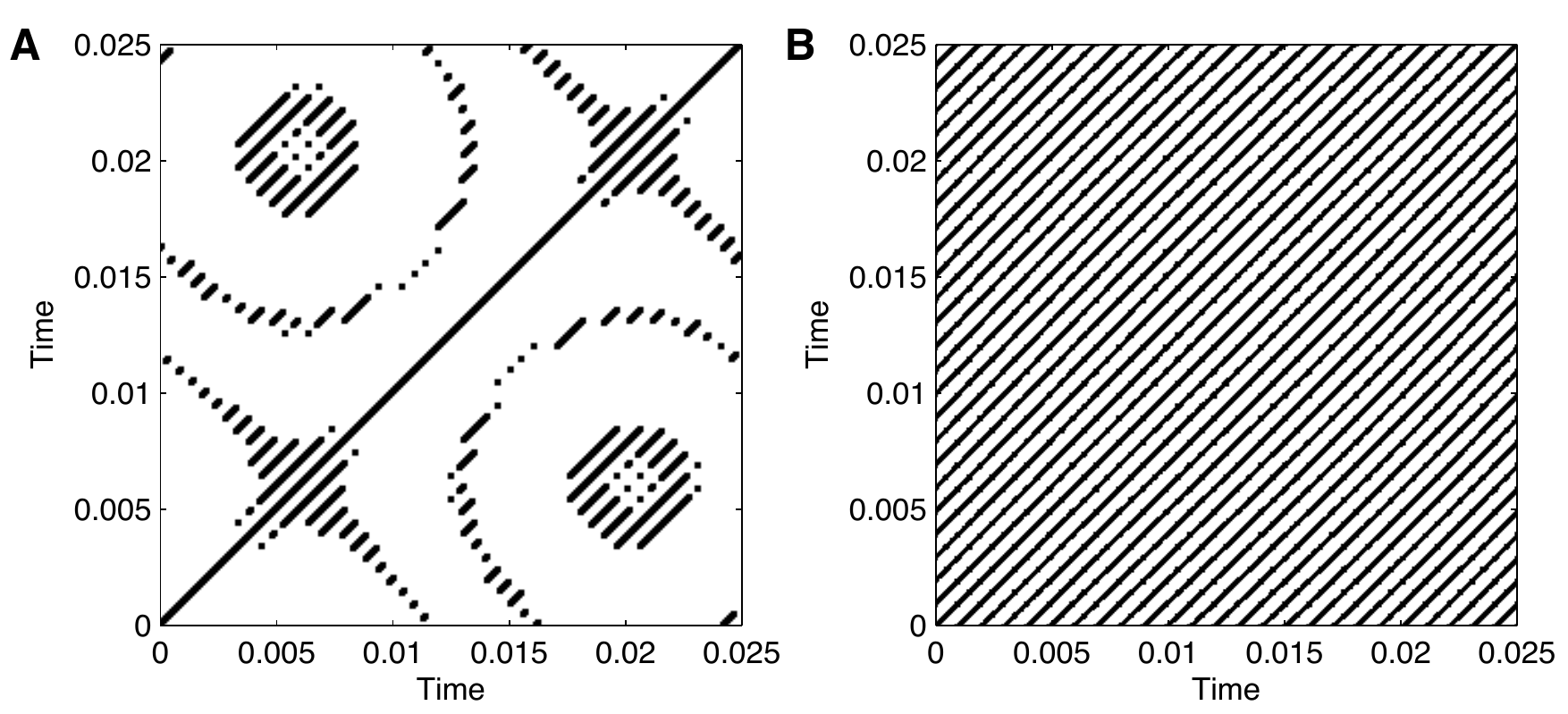}
\caption{(A) Pattern of gaps (all white areas) in a recurrence 
plot of a modulated harmonic oscillation 
$\cos(2 \pi 1000t + 0.5 \sin(2\pi 38 t))$ sampled with 1~kHz.
These gaps represent missing recurrences due to the sampling
frequency close to the frequency of the harmonic signal.
(B) Corresponding RP as shown in (A), but for a higher sampling rate
of 10~kHz. As expected, the entire RP now consists of the
periodic line structures due to the oscillation.
Used RP parameters: $m=3$, $\tau=1$, $\varepsilon=0.05\sigma$, 
$L_{\infty}$-norm.
}
\label{gaps_rp}
\end{figure}
As we can see, a main part of theoretical and methodical work
was now done by the group of Marwan, Romano and Thiel in
Potsdam. Consequently, a first international workshop\index{recurrence plot|workshop} exclusively
dedicated to recurrence plots was organised in 2005 in Potsdam,
Germany (33 participants).

Instead of using spatial information of the phase space trajectory
for the definition of recurrence, Groth has suggested 
to use the local rank order\index{order patterns recurrence plot} \cite{groth2005}. The
local rank order defines specific order patterns whose
recurrences are represented by the {\it order
patterns RP}. This definition of a RP
can help to overcome problems with changing amplitudes (e.g.~drift).

The work of the Potsdam group was continued by Zou,
Ngamga, and Schinkel who worked on a theoretical approach for recurrences
of quasiperiodic systems\index{quasiperiodic system} \cite{zou2007b,zou2007a}, 
on different kinds of transitions, as to strange non-chaotic
attractors\index{strange non-chaotic
attractors} \cite{ngamga2007}, and on order patterns RPs \cite{schinkel2007}.

The sampling rate\index{sampling rate|influence,recurrence plot|artifacts} of oscillating signals can be of importance
for the detection of recurrences \cite{facchini2007a,facchini2005}.
Under certain conditions, large gaps can appear in a RP
where actually recurrence points should be (Fig.~\ref{gaps_rp}). This feigned 
disadvantage can be indeed rather helpful for the detection
of slight frequency changes in oscillating signals which are 
not visible by standard spectral analysis.

A second international workshop on RPs was organised in 2007,
this time in Siena, Italy (44 participants).\index{recurrence plot|workshop}

In 2008, Rohde et al.~linked statistical properties of the
distance matrix to the variance and covariance (at least
for stochastic processes) \cite{rohde2008}. Krishnan
et al.~considered RPs from a completely different point
\cite{krishnan2008a,krishnan2008b}. They stressed the fact that a RP
can be considered as the adjacency matrix of a complex
network\index{complex network}, allowing topological analysis of networks
or graphs by means of RQA. This approach is especially interesting
in many interdisciplinary scientific research.

\section{The spreading application fields}
\label{sec6}

In the last years, RPs again received more attention. Since
2005, more than 50 publications\index{recurrence plot|publications} appear per year (Fig.~\ref{publications}).
Whereas in the beginning of the applications of RPs, the
method was mainly applied in life sciences (e.g.~cardiology,
neuro-psychology),
the method became popular in other scientific fields
during the years. Starting in 1994, a first application\index{recurrence plot|applications}
in earth sciences \cite{kurths94}, in 1996 in finance \cite{gilmore96},
and in  1999 in engineering 
\cite{elwakil99}, chemistry \cite{rustici99} and applied physics \cite{vretenar99}
appeared. Since 2000 we find numerous applications in many disciplines,
from physiology, to biology, earth sciences, acoustics, engineering and
material sciences, finance and economics,
to fundamental research in chemistry and physics (for examples
we refer to \cite{marwan2007}). 

\begin{figure}[htb]
\centering \includegraphics[width=\columnwidth]{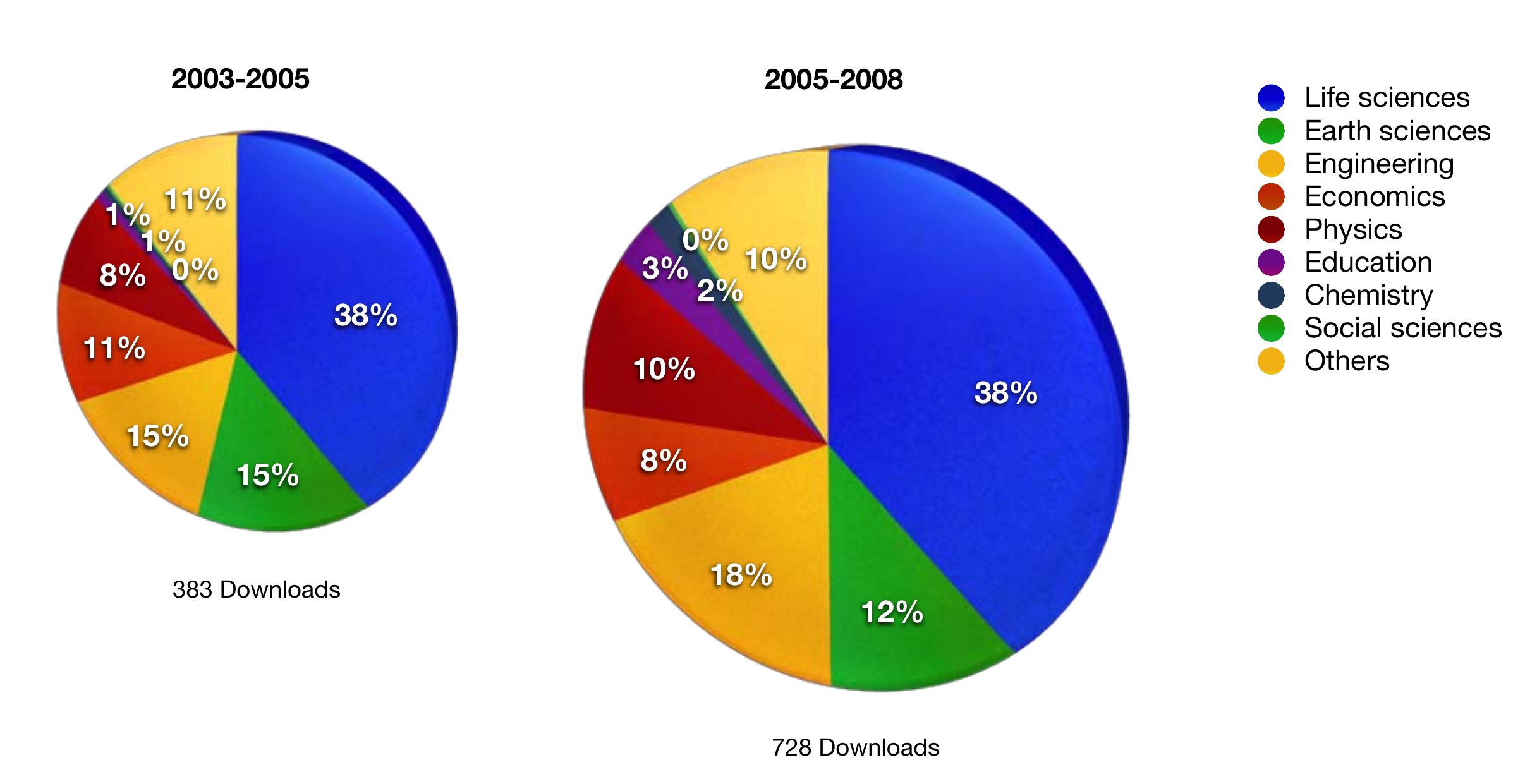}
\caption{Usage of the CRP Toolbox for Matlab since 2003. 
}
\label{toolbox_usage}
\end{figure}

For the usage of the CRP Toolbox (cf.~\url{http://www.recurrence-plot.tk})
we have a statistic of the main purposes of application, which
allows us to estimate the distribution of applications of recurrence
plot based techniques in different scientific disciplines
since 2003 (Tab.~\ref{tab_toolbox_usage}, Fig.~\ref{toolbox_usage}). 
Although we found few repetitions of downloads, the main distribution
of application fields is not affected by such repetitions. A further
problem in analysing these data is, that we sometimes got multiple 
choices of scientific fields, even rather unlike combinations,
like earth science and neuro science. The selection of the scientific fields 
and sub-fields may occur rather arbitrary. We do not claim that it
is a complete and best selection. However, it is mainly based on the
submitted scientific fields or research interests of the users. Some
noteworthy and interesting fields are hidden within the more general
subjects, like artificial intelligence (in engineering), image processing
or telecommunications (in computer and IT networks) or volcanology
(in geophysics). Several users have not provided information about
the intended purpose. We should
also mention that we ensure a high data policy and use the provided data only for
a statistical analysis like this.

For a usage statistics we consider two separate periods: 
a first period between May 2003 and October 2005
with 383 downloads and a second between November 2005 and May 2008 with 728
downloads, revealing the increasing popularity of RPs and the needs of a
corresponding Matlab toolbox. The distribution of the application fields
has only slightly changed between these two periods; only the increase
of applications in engineering (from 15\% to 18\%) 
and the slight decrease in earth sciences (from 15\% to 12\%)
is remarkable (Fig.~\ref{toolbox_usage}). Therefore, in the following
we discuss only the second period. The main application fields are life 
sciences (275 downloads), where psychology, neuro and cognitive sciences (EEG measurements) take the largest part (152 downloads) and cardiology only the third largest part (36 downloads) behind 
different medical problems (75 downloads). The next application
fields are engineering (131), earth sciences (89),
physics (72), economics (55), education (21), chemistry (12) and
even social sciences (2). For 71 downloads we have not received 
sufficient information about the purpose of the usage.

\begin{table}[htb]
\begin{center}
\caption{Scientific fields of usage of the CRP Toolbox since 2003 (May 2003 -- October 2005,
November 2005 -- May 2008; descending order of usage in period 2005--2008).}\label{tab_toolbox_usage}

\begin{tabular}{llrrrr}
\hline
Field			&Subject										&\multicolumn{2}{c}{2003--2005}	&\multicolumn{2}{c}{2005--2008}\\
\hline
\hline
Life sciences	&Psychology/ cognitive and neuro sciences		&54		&147	&152	&275\\
				&Medicinal research/ bio-electronics				&59		&		&75		&\\
				&Cardiology										&24		&		&36		&\\
				&Genomics/ DNA sequencing						&2		&		&6		&\\	
				&Proteins/ systems biology						&8		&		&6		&\\
\hline
Engineering		&Engineering									&39		&56		&63		&131\\	
				&Computer and IT networks						&7		&		&30		&\\	
				&Speech signals/ audio analysis					&8		&		&18		&\\
				&Traffic and transportation						&2		&		&14		&\\
				&Metal processing and analysis					&0		&		&6		&\\	
\hline
Earth sciences	&Atmosphere and weather/ climatology			&10		&59		&29		&89\\
				&Solar and astrophysics							&9		&		&14		&\\	
				&Hydrology										&4		&		&12		&\\	
				&Ecology										&19		&		&11		&\\	
				&Geology										&5		&		&9		&\\
				&Geophysics										&4		&		&5		&\\	
				&Seismology										&6		&		&4		&\\	
				&Geography										&2		&		&5		&\\	
\hline
Physics			&Applied physics								&20		&31		&38		&72\\
				&Theoretical physics							&11		&		&34		&\\	
\hline
Economics		&Finance and markets							&35		&41		&41		&55\\	
				&Economics										&6		&		&14		&\\
\hline
Education/ Teaching		&										&		&3		&		&21\\
\hline
Chemistry		&												&		&2		&		&12\\
\hline
Social sciences	&												&		&1		&		&2\\
Others			&												&		&43		&		&71\\
\hline
\end{tabular}
\end{center}
\end{table}

\section{Outlook}
\label{sec7}
A rather curious sign that RPs are at the final step to really become 
widely known and accepted, we conclude with the 2008 April hoax of
the Australian office of the internet company Google. In a press
release on April 1st, 2008, Google announced the launch of 
a new search technology called {\it gDay}, which would be
able to accurately predict future internet content\index{recurrence plot|Google} \cite{google2008}:

\begin{quote}
``\ldots Using Google's index of historic, cached web content 
and a mashup of numerous factors including recurrence plots 
and fuzzy measure analysis, gDay creates a sophisticated 
model of what the internet will look like 24 hours from now -- 
including share price movements, sports results and news events.
\ldots ''
\end{quote}

As we know, many things Google introduced turned out to be quite popular later.

\section*{Acknowledgement}
The author thanks A.~Giuliani, C.~Webber~Jr.~and J.~P.~Zbilut for 
helpful comments and suggestions.
This work has been supported by the project MAP AO-99-030 
of the Microgravity Application Program/ Biotechnology from the 
Human Spaceflight Program of the European Space Agency (ESA).
\newpage
\bibliography{rp,mybibs,others}
\bibliographystyle{plain}
\end{document}